\documentclass[letterpaper, 10 pt, conference]{ieeeconf}  % Comment this line out if you need a4paper

\IEEEoverridecommandlockouts                              % This command is only needed if 
 \usepackage[bookmarks=true]{hyperref}
\usepackage{multicol}
\usepackage[ruled,longend]{algorithm2e}
\usepackage[bookmarks=true]{hyperref}
\RequirePackage{etex} 
\usepackage{graphics} %
\usepackage{subfigure}
\usepackage{epsfig} %
\usepackage{mathptmx} %
\usepackage{times} %
\usepackage{amsmath} %
\usepackage{amssymb, amsfonts}  %
\usepackage{multirow}
\usepackage{color}
\usepackage{amsfonts}
\usepackage{xcolor}
\usepackage{tikz}
\usepackage{etex}
\usepackage{arydshln}
\usepackage{footnote}
\usepackage{floatrow}
\usepackage{booktabs}
\usepackage{inputenc}
\usepackage{wrapfig}
\usepackage{setspace}

\usepackage{bibentry}

\newcommand{\lenactions}{11 actions} %This number keeps changing

\newcommand{\threshold}{3 meters}
\title{\LARGE \bf
Fast Decision Support for Air Traffic Management at Urban Air Mobility Vertiports using Graph Learning
}

\author{Prajit KrisshnaKumar$^{1*}$, Jhoel Witter$^{2*}$, Steve Paul$^{3*}$, Hanvit Cho$^{4*}$, Karthik Dantu$^{5\ddagger}$, and Souma Chowdhury$^{6*\dagger}$% 
\thanks{$^\dagger$ Corresponding Author, soumacho@buffalo.edu}
\thanks{$^{*}$Department of Mechanical and Aerospace Engineering,
        University at Buffalo, Buffalo, NY 
        {\tt\small \{prajitkr, jhoelwit,stevepau,hanvitch\}@buffalo.edu}}%
\thanks{$^{\ddagger}$Department of Computer Science and Engineering,
        University at Buffalo, Buffalo, NY
      {\tt\small kdantu@buffalo.edu}  }%
\thanks{This work was supported by Stephen Still Institute for Sustainable Transportation and Logistics (SSISTL) at University at Buffalo and National Science Foundation (NSF) award CMMI 2048020. Any opinions, findings, conclusions, or recommendations expressed in this paper are those of the authors and do not necessarily reflect the views of the SSISTL and NSF.}% <-this % stops a space
\thanks {\copyright\space2023 IEEE. Personal use of this material is permitted. Permission from IEEE must be obtained for all other uses, in any current or future media, including reprinting/republishing this material for advertising or promotional purposes, creating new collective works, for resale or redistribution to servers or lists, or reuse of any copyrighted component of this work in other works.}
}

\begin{document}

\maketitle
\thispagestyle{empty}
\pagestyle{empty}

%%%%%%%%%%%%%%%%%%%%%%%%%%%%%%%%%%%%%%%%%%%%%%%%%%%%%%%%%%%%%%%%%%%%%%%%%%%%%%%%
\begin{abstract}

Urban Air Mobility (UAM) promises a new dimension to decongested, safe, and fast travel in urban and suburban hubs. These UAM aircraft are conceived to operate from small airports called vertiports each comprising multiple take-off/landing and battery-recharging spots. Since they might be situated in dense urban areas and need to handle many aircraft landings and take-offs each hour, managing this schedule in real-time becomes challenging for a traditional air-traffic controller but instead calls for an automated solution. This paper provides a novel approach to this problem of Urban Air Mobility - Vertiport Schedule Management (UAM-VSM), which leverages graph reinforcement learning to generate decision-support policies. Here the designated physical spots within the vertiport's airspace and the vehicles being managed are represented as two separate graphs, with feature extraction performed through a graph convolutional network (GCN). Extracted features are passed onto perceptron layers to decide actions such as continue to hover or cruise, continue idling or take-off, or land on an allocated vertiport spot. Performance is measured based on delays, safety (no. of collisions) and battery consumption. Through realistic simulations in AirSim applied to scaled down multi-rotor vehicles, our results demonstrate the suitability of using graph reinforcement learning to solve the UAM-VSM problem and its superiority to basic reinforcement learning (with graph embeddings) or random choice baselines. 

\end{abstract}

\section{INTRODUCTION}
% Urban Air Mobility (UAM) transforms the realm of transportation by exploiting air travel with Vertical Takeoff and Landing (VTOL) vehicles. Developments in this area can lead to implementations such as air taxis \cite{rothfeld2020urban} \cite{BAURANOV2021100726} \cite{horne2019next}, or even
\vspace{-1mm}
%With an estimated 68\% of the world population residing in urban areas \cite{united_nations_2018} and more than 10\% increase in average commute time a new mode of transportation is inevitable. 
Urban Air Mobility (UAM) based on electric Vertical Takeoff and Landing (eVTOL) aircraft is becoming an increasingly popular concept for future urban transportation, with the objective to mitigate problems such as traffic congestion and circuitous routes. Such eVTOLs are electric aircraft with the ability to transport 2-5 passengers, air-ambulance or emergency services, or serve in a goods transport role. Over the past few years, there have been numerous feasibility studies on UAM in the context of addressing traffic congestion, and its overall impact on reducing carbon emissions \cite{kadhiresan2019conceptual,goyal2018urban,chao2021weather}. 
% One of the major problems with current urban/suburban transportation is the increase in traffic congestion and circuitous routes, especially in major metropolitan areas. With the turn of this millennium, there has been a steady increase in the average commute from 25 minutes in 2006 to 27.6 min in 2019, which is a 10$\%$ increase as discussed in \cite{ACS}. This issue is further exaggerated by the continuing  migration of the population from rural to urban areas. By the year 2050, almost 68\% of the world's population will reside in urban areas \cite{united_nations_2018}. In order to address the above-stated problem, one approach is to introduce Urban Air Mobility (UAM) with electric Vertical Takeoff \& Landing (eVTOL) aircraft as a new dimension in general transportation. 
% As a result, there has been significant interest from various government agencies and private organizations in the form of funding and investment over the last decade \cite{21,straubinger2020overview} in this technology. According to \cite{UAMmarket}, by the year 2030 the UAM market can be a 9.1 billion dollar industry. 
% 
However, the realization of UAM comes with its own set of challenges which includes the design challenges with regards to vertiport infrastructure, accessibility and noise impact, and operational challenges which include UAM fleet scheduling \cite{Paul_ICRA} and vertiport operation \cite{vertiporthigh}. In these contexts, safety is the primary driving factor behind the planning processes \cite{vertiporthigh}. A Vertiport or aerodrome is conceived as a site where the eVTOLs take off, land, and charge their batteries \cite{daskilewicz2018progress}. In this work, we focus on the problem of vertiport operation, specifically real-time scheduling of eVTOL take-off and landing. The key objectives are to ensure safety (i.e., avoid collisions), minimize delays with respect to a pre-defined schedule of takeoff and landing, and conserve battery. This problem is named UAM-Vertiport Schedule Management (UAM-VSM). 
% {\color{blue}End with :} Why vertiport operation is important?
% 	{\color{blue}Difficulty of operation management. How is it different from train, subway or flight}

Once deployed, such UAM vehicles are expected to be entering or leaving a vertiport located in a dense urban environment \cite{guerreiro2020capacity}, somewhat similar to a subway network in major cities, and unlike typical airports used in general aviation. Unlike in ground transportation, minor collisions or undesirable proximity can lead to dire consequences, thereby necessitating strict spatiotemporal constraints on their operation. Assuming high traffic at any given vertiport (which is expected for an economically viable UAM network) and uncertainties due to weather and system failures, over-reliance on a human Air Traffic Controller (ATC) to perform the real-time vertiport traffic management is unduly risky. The high frequency of decision-making could lead to a cognitive overload for the ATC and result in poor decision-making \cite{mathur2019paths}. Hence, there is an urgent need to develop \textit{automated} approaches to manage the real-time scheduling of takeoff/landing of eVTOL aircraft at vertiports.
% Considering the pace with which UAM is becoming a reality and the challenges to be overcome to ensure safe and efficient operation, there is an urgent need to develop \textit{automated} approaches to takeoff/landing scheduling at vertiports.
  
Prior works on UAM scheduling have mostly focused on eVTOL fleet scheduling and path planning across multiple vertiports in a region \cite{9832471, Paul_AVIATION_2022} to maximize profit and meet potential travel demand. Very limited quantitative work exists in automated decision-support for real-time scheduling of takeoff/landing with the objectives to maximize safety and minimize delays. Unlike the other works which schedule higher-level decisions such as which destination/vertiport to visit next, or when to charge an eVTOL \cite{simpson1968computerized}, here we assume that a high-level schedule over a longer time horizon of say 6-12 hours already exists. Our focus is on computing real-time decisions during each 1-minute time window, during which tasks (land, takeoff, move to spot, etc.) have to be allocated to up to 4 aircraft within the operational air space managed by that vertiport. In addition, unlike prior works where the eVTOL is simulated by a simple linear model \cite{theron2020integrated}, we consider a more realistic eVTOL simulation developed in AirSim \cite{shah2017airsim}, based on a scaled-down multi-rotor aerial vehicle model. This is done to ensure that the simulated outcomes have a reduced reality gap. Moreover, this work is among the first to also present a scaled-down physical validation of the proposed solution to UAM-VSM, through indoor experiments with four small quadcopters. 

Due to high computational costs, traditional methods such as Mixed Integer Non-Linear Programming (MINLP) and heuristic-search-based methods are unviable for the real-time decision-making needs of the UAM-VSM problem. Hence a learning-based method is posited as a robust alternative, which can yield policies that are real-time executable. Specifically, we take a novel graph-learning approach to solving this problem in a scalable and generalizable manner. To do so, firstly we consider the different designated physical spots (\autoref{figs:state-flow}) that can be allocated to a vehicle, such as landing/take-off pad (or spot), hovering spot and charging spot or battery port, as nodes in a graph. We also consider the eVTOLs aircraft to be represented as another \textit{Graph}. With this representation, we exploit Graph Neural Network (GNN) for feature extraction over the graph-encoded state-space, and Proximal Policy Optimization (PPO) \cite{schulman2017proximal} is used to train this policy network. 

 \textbf{Key Contributions:} The primary contributions of this work lie in: 1) Formulating the UAM-VSM problem as a Markov Decision Process (MDP) and development of a relatively realistic virtual environment to simulate this MDP for scaled-down multi-rotor vehicles; 2) Testing the hypothesis that a graph abstraction of the state-space of designated locations in the vertiport's airspace and vehicles being managed, along with graph learning, can yield effective policies for this problem; 3) Analyzing the trade-offs between objectives such as takeoff delay, landing delay, collision avoidance, and battery conservation, achieved with such graph-learning based UAM-VSM policies. Contribution 2 is demonstrated via comparisons against vanilla RL solutions, randomized decision-making, and first-come-first-serve decision-making, with all methods evaluated over a sampling of different 24-hour operational periods.  
 %In this paper, the terms \textbf{eVTOL}, \textbf{VTOL}, \textbf{drone}, \textbf{vehicle} and \textbf{robot} refer to the same entity, i.e., a UAM aircraft.

The rest of this paper is structured as follows. In Sec. \ref{sec:Related_works}, some of the prior works related to the problem is presented. Sec. \ref{sec:prob_formulation} describes the formulation of UAM-VSM as an MDP, followed by the description of the proposed state abstraction and graph learning architecture, in Sec. \ref{sec:learning}. In Sec. \ref{sec:sim_env}, we present the simulation environment developed for Vertiport management problems. Sec. \ref{sec:results} presents the numerical experiments and comparison of results with that of baselines, namely a random method, a first-come-first-served method, and a standard RL approach. Sec. \ref{sec:conclusion} summarizes our concluding remarks.

\section{Related work}\label{sec:Related_works}
\vspace{-2.5mm}
In commercial aviation, the most common approach for aircraft takeoff/landing schedule is the First-Come-First-Served (FCFS) approach \cite{ODONI1994107}. The main advantage of this approach is that it is easy to implement without the need for any sophisticated scheduling software. The only constraint for this approach is maintaining a safe distance. However, the major drawback of this approach is that it does not provide any scope for improvement to decrease time delays. Trivizas and Lieder et al. \cite{trivizas1998optimal,RePEc:eee:ejores:v:243:y:2015:i:1:p:61-69} proposed a Dynamic Programming algorithm for optimal landing on a runway, while Beaslet et al. and Abela et al. \cite{10.1287/trsc.34.2.180.12302,Abela1993ComputingOS} proposed a Mixed Integer Programming (MIP). These approaches are feasible only for commercial aviation takeoff/landing where the number of schedules is of the order of 1 every few minutes. In recent years, RL algorithms with Graph Neural Networks or GNN are being increasingly used to solve planning and scheduling problems such as TSP, VRP, Max-Cut, Min-Vertex, and MRTA~\cite{khalil2017learning, 9750805, Paul_ICRA, li2018combinatorial, paul2022scalable, nowak2017note,9750805, Paul_AVIATION_2022, paul2023efficient}. Some of the recent works \cite{Paul_AVIATION_2022,shihab2020optimal,9832471, 9482700} on UAM mainly focus on scheduling routes between vertiports with the aim to optimize an entity such as profit, delay, and demand met. 
Here, we build on our prior work on this under-addressed problem \cite{kumar2023graph}, by considering the critical objective of safety and presenting comparisons with a standard RL implementation that does not use graph abstractions.

% {\color{blue} Explain design choices: 1) why the graph formulation, 2) Why the GNN and specifically GCN, 3) Why PPO?}

 \vspace{-.2cm}
\section{Problem Formulation}
\label{sec:prob_formulation}
\begin{figure}
\centering
\scriptsize
\includegraphics[width=0.9\linewidth]{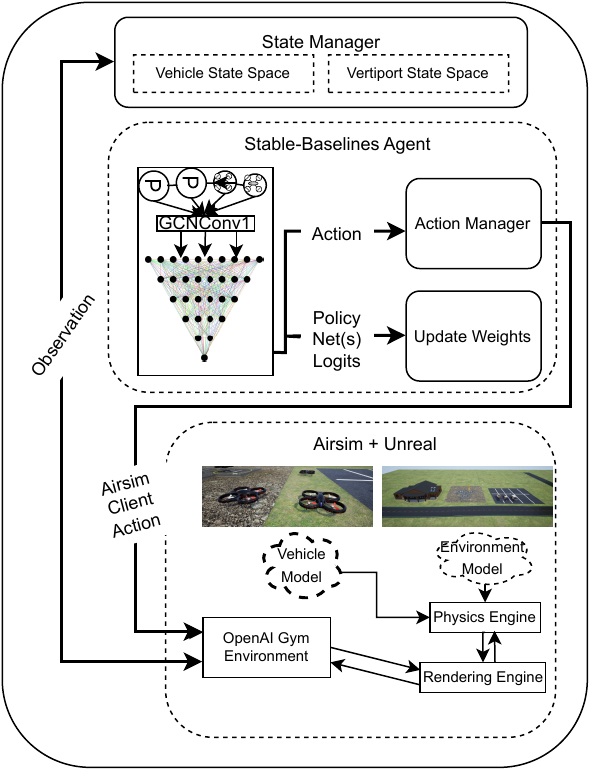}
\caption{Overview of Simulation Environment displaying the progression of Information through the State Manager, Action Manager, Learning Policy Network, and the Simulation Engine (Airsim and Unreal Engine)}
\label{figs:flow}
\end{figure}
\begin{figure}
\centering
\scriptsize
\includegraphics[width=0.5\linewidth]{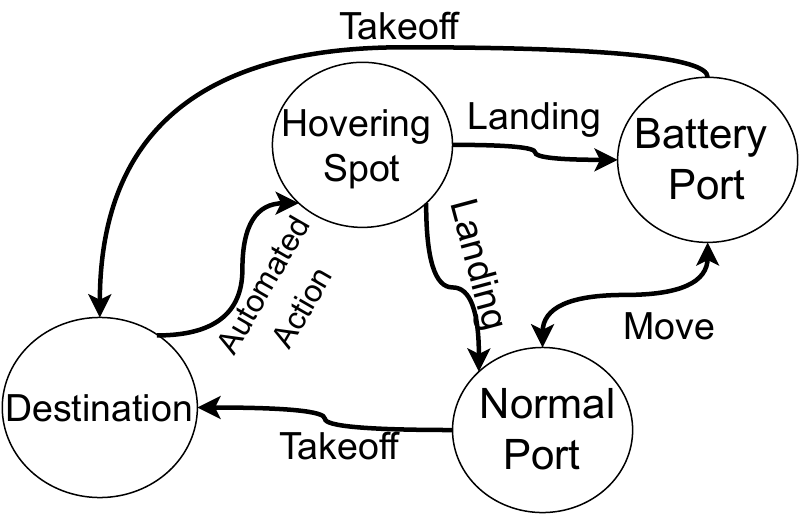}
\caption{State-Transition Diagram}
\vspace{-2mm}
\label{figs:state-flow}
\end{figure}

We consider this an Urban Air Mobility-Vertiport Schedule Management (UAM-VSM) problem, where the goal is to design an automated centralized Air Traffic Controller (ATC) agent in order to successfully: {\bf i.)} assign tasks to incoming vehicles (taking off to destinations, landing, hovering); {\bf ii.)} maintain a sufficient battery level among all vehicles being controlled; {\bf iii.)} ensure a safe flight for each vehicle without collisions. Throughout this paper, we assume there is no communication loss and no uncertainties in the environment. To train the agent, we created an environment identical to that of a realistic vertiport, with two ports for landing (Normal Ports), and one charging station (Battery Port). There are four hovering spots arranged around the ports, as well as five destinations away from the agent's airspace that the vehicles travel to and from. The overall framework is shown in Figure \ref{figs:flow}. Each vehicle has 4 states and the state-transition diagram is shown in figure \ref{figs:state-flow}. 

%Environment stuff
% The environment is initiated with the number of eVTOLs and the type of learning architecture (either graph learning or baseline learning). This creates the Airsim client which is used for all interactions with the simulated Unreal Engine (UE) environment \cite{shah2017airsim}, \cite{sanders2016introduction}, as well as the action and observation space for the reinforcement learning algorithm in use. Each eVTOL starts at a hovering spot or at a destination, with set takeoff and landing scheduled times respectively. For training, each scheduled time for takeoff is issued between 10 and 20 minutes after the eVTOL returns to the Vertiport, and likewise the landing time is issued between 5 and 15 minutes after the eVTOL begins traveling back to the Vertiport from a destination. 

The environment is initiated with 4 vehicles at a random state and with a random schedule. Takeoff time is issued between 10 and 20 minutes after the vehicle returns to the vertiport, and likewise, the landing time is issued between 5 and 15 minutes after the vehicle begins traveling back to the vertiport from a destination. Each vehicle starts with a full battery and discharges at a rate of $discharge\_rate$ ($\Omega$) per step. $\Omega$ is calculated based on the action:
\begin{equation}
\footnotesize
\Omega =
\left\{
	\begin{array}{ll}
		d_t  & \mbox{if cruising}\\
        0.5  & \mbox{if hovering}\\
		0.25 &  \mbox{if idling on ground}\\
	\end{array}
\right.
\end{equation}
% \begin{figure*}
% \centering
% \includegraphics[width=0.9\linewidth]{Fig/UAM-takeofflanding-envflow.drawio.png}
% \caption{Simulation Environment Overview Showing the Flow of Information through the State Manager, Action Manager, Learning Policy Network, and the Simulation Engine (Airsim + Unreal)}
% \label{figs:flow}
% \end{figure*}
where $d_t$ is the distance from one point to the next. When a vehicle rests at a charging station, it regains 10\% battery per time-step. Thus a vehicle can lose battery even while not moving, which will encourage the agent to travel and engage with the charging port.

Vehicles move independently, allowing multiple simultaneous movements. This setup helps the agent learn collision avoidance when vehicles intersect. At each time step, the agent selects an available vehicle at the port to act. If the chosen vehicle has reached its destination, the agent waits until it re-enters the vertiport airspace before acting. The simulation runs 300 times faster than in real-time, so every second in the simulator corresponds to 5 minutes in real-time and is useful for training, as it allows for step times as low as 0.2 seconds, or  ~288 seconds (4.8 minutes) per episode (1440 steps). Every episode simulates a full day of operation. Every vehicle is updated with a minimum frequency of $45 Hz$, which includes updating all features (location, delay, battery status, and vehicle status). 

% \begin{algorithm}
% % \vspace{3mm}

%   $I$: Total Number of VTOLs;
%   $G$: Graph\;
%   $S \gets$ Skeletonize image I\;
%   $G.\texttt{init}(x_\text{init})$\;
%   \For{\texttt{vehicle} in \texttt{medial-axis of} S}
%   {
%     $x_\text{new} \gets \text{\texttt{pixel position}}$\;
%     $x_\text{edge} \gets Edge(x_\text{init},\ x_\text{new})$\;
%     $G.\texttt{addnode}(x_{\text{new}})$,\ $G.\texttt{addedge}(x_{\text{edge}})$\;
%     $x_\text{init} \gets x_{\text{new}}$\;
    
%   }
%   \KwRet{$G$}\;
  
%   \caption{The skeleton graph for path planning}
%   \label{algo:path_planning}

% \end{algorithm}

\subsection{MDP Formulation}
\label{sec:MDP}
% The problem is formulated as Markov decision process(MDP), the state, action and reward can be defined based on equation \ref{eq:mdp}

% \begin{equation}\label{eq:MDP} 
%     \begin{aligned}
%      &s = \mathcal{F}_s(N_{C,i},\ \Delta(C_{C,i},X_{k^*}), \ P(G_l),\ t)\\
%      &a = \mathcal{F}_a(N_{S,j},\ X_{k^*_j},\ \gamma_j)\\
%      &r = \mathcal{F}_r(\Delta(C_{C,i},\ G_l), \psi_l,\ N_{C,i},\ t)\\
%     \end{aligned}
% \end{equation}

% \begin{wrapfigure}[7]{r}{0.55\linewidth}
% \vspace{-1.1cm}

\begin{table}
\vspace{-4mm}
\footnotesize
\centering
\begin{tabular}{|l|l|} 
\hline
Type                              & Variable                         \\ 
\hline
\multirow{3}{*}{Vertiport States} & Availability - $P_a$                     \\
                                  & Port type - $P_t$                  \\
                                  & Location - $(x_p, y_p)$                     \\ 
\hline
\multirow{4}{*}{VTOL States}     & Current status - $c_i$                   \\
                                  & Battery capacity - $b_i$                \\
                                  & Schedule status - $l_i$                 \\
                                  & Location - $(x_i, y_i)$                         \\
                                 
\hline
\multirow{5}{*}{Action Space}     & Stay still                       \\
                                  & Takeoff                          \\
                                  & Move/ land in normal port - 1,2  \\
                                  & Move/ land in battery port - 1   \\
                                  & Move to hover spots - 1,2,3,4    \\
                                  & Continue previous action        \\
                                  & Avoid collision                 \\
\hline
\end{tabular}
\caption{MDP formulation}
\label{tab:MDP_test}
\vspace{-4mm}
\end{table}

The Urban Air Mobility-Vertiport Schedule Management problem is modeled as a Markov Decision Process (MDP), where the state space consists of the state of the vertiports and the state of the vehicles. 
The vertiport state information is represented as graph $\mathcal{G}_{VP}$=$(V_{VP}, E_{VP})$, where $V_{VP}$ represents the vertices or nodes of the graph and $E_{VP}$ represents the set of edges. Every node $i \in V_{VP}$ has the following properties (also described in table \ref{tab:MDP_test}) $\delta^{VP}_{i}$ = [$P^{i}_a$, $P^{i}_t$, $x_p$, $y_p$].
Similarly, the vehicle state information is represented as a graph $\mathcal{G}_{EV}$=$(V_{EV}, E_{EV})$, where $V_{EV}$ represents the nodes of the graph and $E_{EV}$ represents the set of edges. Every node $i \in V_{EV}$ has the following properties (also described in table \ref{tab:MDP_test}) $\delta^{EV}_{i}$ = [$c_i$, $b_i$, $l_i$, $x_i, y_i$]. 
While there are no explicit environment uncertainties here, state transition in principle is not deterministic due to the possibility of aircraft collision (that triggers avoidance actions) modeled by the simulation. 

% Therefore the state space consists of \textbf{1)} The Vertiport graph $\mathcal{G}_{VP}$, \textbf{2)} The EVTOL graph $\mathcal{G}_{EV}$, \textbf{3)} ...., \textbf{4)}...., \textbf{5)}....  

% , action, and reward is given below:
% \vspace{-2mm}
% \begin{equation}\label{eq:MDP} 
%     \begin{aligned}
%      &s = [b_{i}, c_{i}, l_{i}, x_{i, p} , y_{i, p}, p_{a, t} ]\\
%      &a = \mathcal{F}_a(s)\\
%      &r = \mathcal{F}_r(a, s, \S, \beta, \gamma, \tau, \gamma, w_n)\\
%     \end{aligned} \vspace{-2mm}
% \end{equation}

% where $i$ and $p$ stand for all the vehicles and ports respectively. 
The state and action space can be found in table \ref{tab:MDP_test}, and the reward is shown in equation \ref{eq:reward}. Here $\tau$ is the takeoff coefficient, $\gamma$ is the landing coefficient, $\lambda$ is the battery coefficient, $\beta$ is the delay coefficient, $\S$ is the safety coefficient, and $w_n$ are the weights. 
\vspace{-.2cm}
\begin{equation}\label{eq:reward} 
\vspace{-3mm}
    \begin{aligned}
        R = w_1\tau + w_2\gamma + w_3\lambda + w_4\beta + w_5\S
    \end{aligned}  \vspace{-3mm}
\end{equation}\\
 \vspace{-.3cm}
Here are more details about the reward terms:\\

\subsubsection{Takeoff-Landing coefficient}
We classify a ''\textbf{good}'' takeoff as one where the vehicle is: {\bf i)} departing punctually (within 5 minutes of its planned takeoff time); {\bf ii)} taking off with a battery level exceeding 30\%. The same standards apply to a ''\textbf{good}'' landing, with the added option that the vehicle can decide to land before its scheduled time. Both $\tau$ and $\gamma$ fall within the range of ${-5,5}$.

\subsubsection{Battery coefficient}
This coefficient is defined as:
\begin{equation}
\footnotesize
\lambda = 
\left\{
	\begin{array}{ll}
		5\times \frac{b_i}{100} & \mbox{if } b_i \geq 30\\
        -5  & else\\
	\end{array}
\right. \vspace{-1mm}
\end{equation}

The value gradually rises as the vehicle charges. To discourage operations at critical battery levels, a penalty is imposed once the battery capacity falls below 30\%. In order to maintain the battery coefficient, the agent must ensure each vehicle is fully charged.

\subsubsection{Delay coefficient}
The delay time is computed from the point when the vehicle misses its window for either taking off or landing. This delay time accumulates till the vehicle receives a new schedule. The delay coefficient is calculated by: \vspace{-.2cm}
\begin{equation}
\vspace{-1mm}
\beta = -5+10 \times e^{-\Delta_i}
\end{equation}\vspace{-.1cm}
where $\Delta$ represents the delay in minutes. This formulation encourages the agent to keep the delay as low as possible by maximizing the delay coefficient.
\vspace{-1.0mm}
\subsubsection{Safety coefficient}
\begin{figure}
% \begin{wrapfigure}[11]{r}{0.7\linewidth}
% \vspace{.2cm}
\centering
\includegraphics[width=0.7\linewidth]{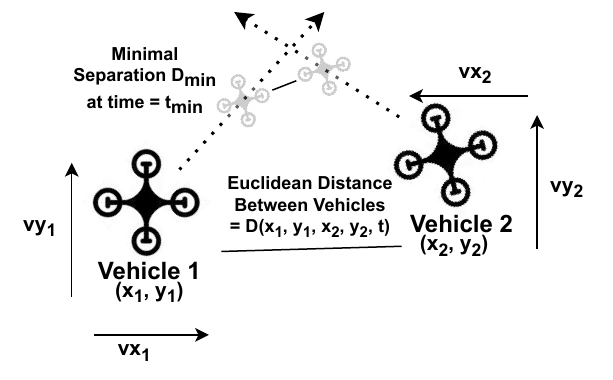}
\vspace{-.4cm}
\caption{Minimal Separation Scenario}
\label{figs:separation}
\end{figure}

% \vspace{-1.85mm}
Before the safety coefficient can be calculated, the environment will check to see if the selected vehicle: {\bf i.)} is currently en-route to a location; {\bf ii.)} has an intersecting path with another vehicle that is en route. This can be visualized in figure \ref{figs:separation}. 

\vspace{-0.5mm}
% \end{wrapfigure}
In the figure, two vehicles are moving toward a common intersection point. The distance between them at any given point is made into a function of time using the Euclidean distance combined with their immediate position and velocity vectors:
\vspace{-.3cm}
\begin{equation}\label{eq:euclid} 
\footnotesize
\vspace{-1.25mm}
D(t) = \sqrt{ (x_1 - x_2 + v_{x_1}t - v_{x_2}t)^2 + (y_1 - y_2 + v_{y_1}t - v_{y_2}t)^2 } 
\end{equation}
where $x_n, y_n, v_{x_n}, v_{y_n}$ are the position and velocity components of vehicles 1 and 2. This equation is then differentiated with respect to time and solved for the local minimum, $t_{min}$:
\vspace{-.4mm}
\begin{equation}\label{eq:tmin} 
\footnotesize
\vspace{-1.25mm}
t_{min} =  -\frac{2(v_{x_1} - v_{x_2})(x_1 - x_2) + 2(v_{y_1} - v_{y_2})(y_1 - y_2)}{2(v_{x_1} - v_{x_2})^2 + 2(v_{y_1} - v_{y_2})^2}
\end{equation}
\vspace{1mm}
$t_{min}$ is then plugged back into equation \ref{eq:euclid} to get the minimum separation $D_{min}$. Each simulated vehicle has an occupant space of 1x1 meters, so if $D_{min}$ is less than \threshold{} and the agent doesn't take evasive action(avoid collision), the agent will be penalized:
\vspace{-.2cm}
\begin{equation}
\footnotesize
\vspace{-1.25mm}
\S =
\left\{
	\begin{array}{ll}
		0  & \mbox{if vehicle is on the ground}\\
		-5 & \mbox{else if } D_{min} \leq 3.0\ \&\ action \neq avoid\ collision\\
		5 & \mbox{else if } D_{min} \leq 3.0\ \&\ action = avoid\ collision\\
	\end{array}
\right.
\end{equation}
% \subsubsection{Reward weights}
% All five coefficients have a range of $[-5,5]$ which brings the range of the combined reward to $[-25,25]$. This is where the weights come in. Each weight $w_n$ is initiated as $\frac{1}{5}$, and tweaked as necessary based on {\bf i.)} The performance of the agent. {\bf ii.)} The importance of environmental factors. In a simple example, imagine the agent is learning how to takeoff and land properly, however it is not avoiding collisions. The weight for the collision coefficient, $w_5$, can be tweaked to give  a higher penalty for collisions, and a higher reward for avoiding them, which in turn can shift the policy into a new local optima. 

\subsubsection{Reward weights}
Each coefficient is multiplied by a weight $w_1, w_2,..w_n$, reflecting the significance of each coefficient. In our scenario, safety is considered the most important, hence the highest weight is allocated to the safety coefficient $\S$, followed by $\beta, \tau, \gamma, \lambda$, in that order.

 \vspace{-.2cm}
\section{Learning Architecture} \vspace{-.2cm}
\label{sec:learning}
This paper focuses on a deep reinforcement learning framework-PPO \cite{schulman2017proximal}- utilizing a GCN for feature abstraction and a Multi-Layered Perceptron (MLP) for prediction. 
\vspace{-2mm}
\subsection{Network Parameters} \vspace{-2mm}
Linear layers with biases are used for all three networks (feature abstraction, value, and policy) except for the feature abstraction network in the GRL agent, for which GCNs are used. We make use of LeakyReLU with a slope of $0.1$ (for feature abstraction and value network) \& Tanh (for the policy network) activation layers, as they help reduce sparse gradients. Adam optimizer with a learning rate of 1e-5 is used for back-propagation. We chose Proximal Policy Optimization (PPO) for the reinforcement learning algorithm \cite{schulman2017proximal}. PPO is based on Trust Region Policy Optimization (TRPO) \cite{https://doi.org/10.48550/arxiv.1502.05477}, and has an objective function tailored to clip policy expansion and allows for a safer policy update.

\vspace{-.2cm}
\subsection{Frameworks} 
\begin{figure} 
\centering
\scriptsize
{\includegraphics[width=0.85\textwidth]{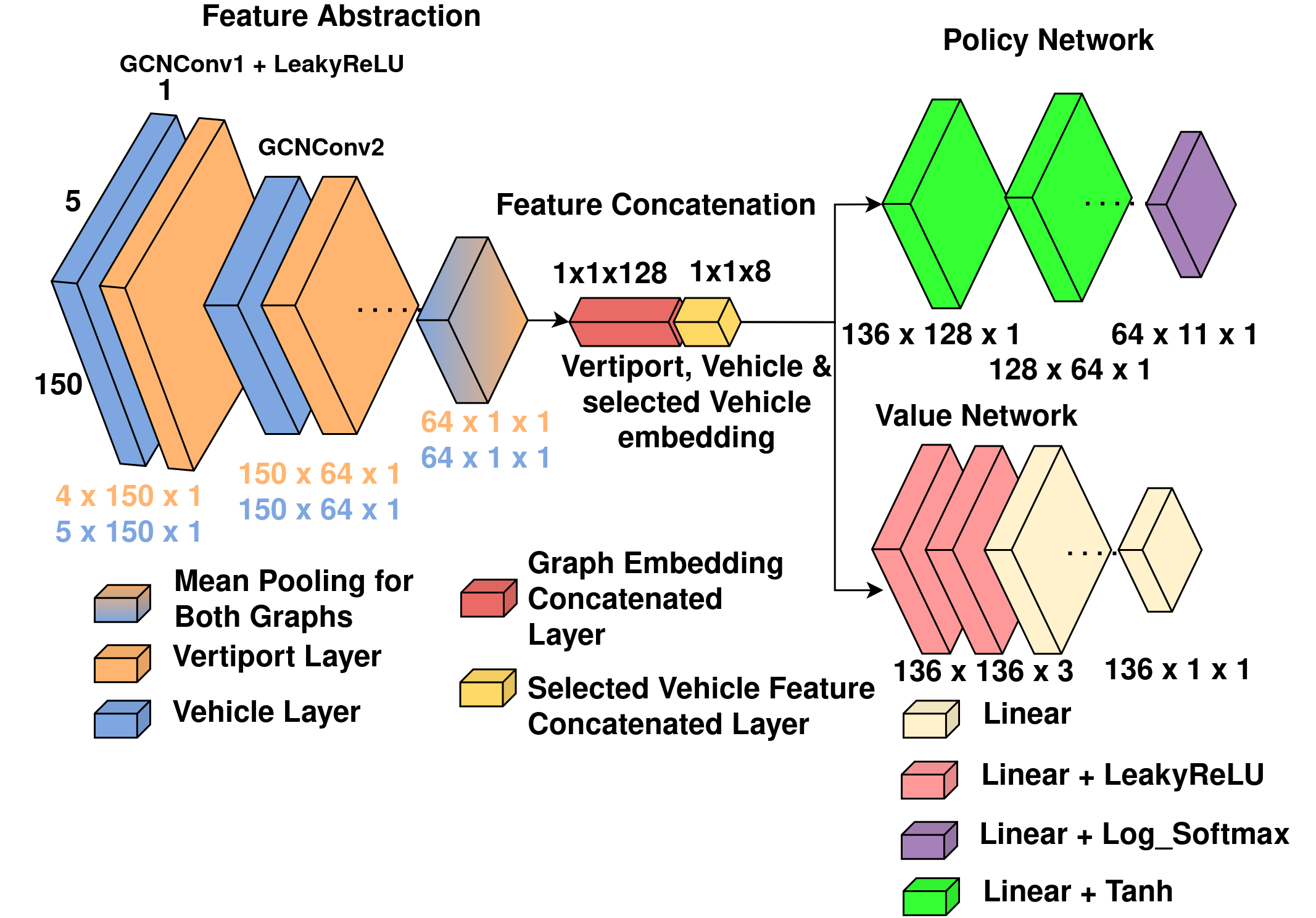}\label{fig:grlagentnetwork_visualize}}
\vspace{-.1cm}
\scriptsize
\caption{The policy network along with the value network for the Graph Learning Architecture}
\label{fig:networkflow_visualize}
\vspace{-5pt}
\end{figure}

  Two frameworks are used for learning the policy model, namely the  Multi-layer perceptron based RL network and the graph learning-based RL network
  Both frameworks have feature abstraction, policy, and value networks to work with PPO. The main difference lies in the feature abstraction network where: the RL agent uses a condensed feature vector with both the vehicle and vertiport state space; the GRL agent uses two GCNs, which take the vehicle and vertiport feature matrix along with their respective edge connectivity matrix. The GRL learning framework is shown in figure  \ref{fig:networkflow_visualize}. Both policy networks will use a four-layer MLP with a log-softmax transformation to obtain log probabilities for the \lenactions{}. Both agents utilize masking which will depend on the state of the selected vehicle and the availability of each port. This takes away a layer of complexity and allows the agent to focus on other environmental factors, such as avoiding collisions and reducing delay.

 \vspace{-.2cm}
\section{Vertiport Simulation}
\label{sec:sim_env}
\vspace{-2mm}
The evolution of the gaming industry over the last decade has had a great impact on the robotics community, allowing us to simulate various complex tasks \cite{behjat2021learning, zeng2022learning}, in a more realistic environment that is nearly impossible in real life due to the hardware limitations. To simulate the Vertiport, a custom 3D environment is developed using Unreal Engine \cite{sanders2016introduction} with AirSim \cite{shah2017airsim} plugin being the backend for physics. AirSim is an open-source simulator plugin developed by Microsoft for unmanned aerial (UAV) and ground vehicles (UGV). Since the operation of VTOLs is similar to that of UAVs, throughout this paper we consider UAVs to be vehicles.  Fig. \ref{figs:flow} shows the overview of the developed simulation environment. Unreal Engine 4 is used as the GUI and AirSim plugin provides dynamics for the vehicle. To reduce the memory and computational footprint of the unreal engine, the entire environment has been created with 101k static triangles, limiting the memory to 3.4 Megabytes(excluding VTOLs). The OpenAI Gym interface provides a connection between Airsim, State/Action Managers, and the RL Agent. This interface receives raw data of all the vehicles from AirSim and filters the necessary data, which then will be transferred to the State Manager. The State Manager and the Action Manager inherit the properties of all the vehicles and ports. Filtering data and structuring the data takes place in the state manager, while the action manager decodes the data from the Policy model to AirSim understandable format.

 \vspace{-.2cm}
\section{Results and Discussion} 
\label{sec:results}
\vspace{-1.25mm}
% \begin{figure}
% \centering
% \includegraphics[width=0.95\linewidth]{Fig/Training/reward.png}
% \caption{Reward (left) and loss (right) for both the GRL and RL agent.}
% \label{fig:reward}
% \end{figure}

% \begin{figure}
% \centering
% \includegraphics[width=0.95\linewidth]{Fig/Testing/metrics.png}
% \caption{Testing metrics for both the GRL and RL agent over 50 episodes.}
% \label{fig:testing}
% \end{figure}

% \begin{figure}
% \centering
% \includegraphics[width=0.95\linewidth]{Fig/Training/collision.png}
% \caption{Collisions for both the reinforcement learning agent and graph reinforcement learning agent.}
% \label{figs:reward}
% \end{figure}

This section presents the results of the case studies performed for the UAM-VSM problem. The following subsections will delve into the {\bf i.)} training analysis of the GRL and RL agent. {\bf ii.)} a comparison of test results with: the GRL agent; the RL agent; a random agent; a First-Come-First-Serve (FCFS) agent. {\bf iii.)} analyses of effects of safety parameter and weights in the reward formulation. Both agents received the same information in their observation space and used the same environment for a fair comparison. Here, we consider only 4 vehicles since the use of realistic simulations limits the number of vehicles that can be simultaneously modeled without blowing up the training process. Further, departing vehicles are not simulated anymore once they leave the port (within their allocated 20-min window), and instead their embodiment is taken over by another aircraft that is either landing or waiting for a take-off schedule, keeping the number of aircraft simulated at 4 at all times.% The training was completed on a core i7 12900H CPU with 64 GB of ram and a RTX 3070Ti GPU. 
% results from the graph learning are compared with the Feasibility-preserving random walk and Reinforcement learning methods. The training is performed in \textbf{pc name} at a clock speed of 300 and the stable baselines 3 \ref{sstable-baselines} is modified based on our configuration for the RL agent.

\vspace{-1mm}
\subsection{Learning Curve}
\begin{figure}
    \centering
    \begin{minipage}[b]{0.5\textwidth}
        \centering
        \includegraphics[width=\linewidth]{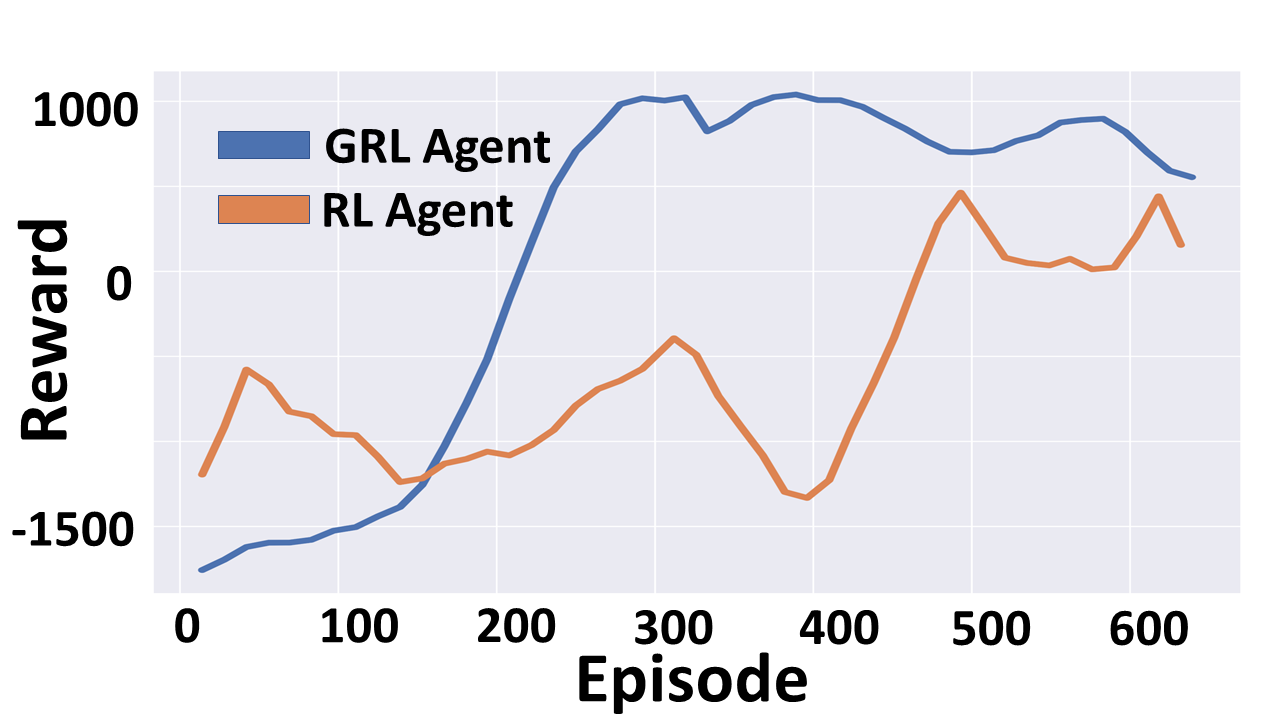}
        % \caption{Reward}
        \label{fig:rlagentnetwork}
    \end{minipage}%
    \begin{minipage}[b]{0.5\textwidth}
        \centering
        \includegraphics[width=\linewidth]{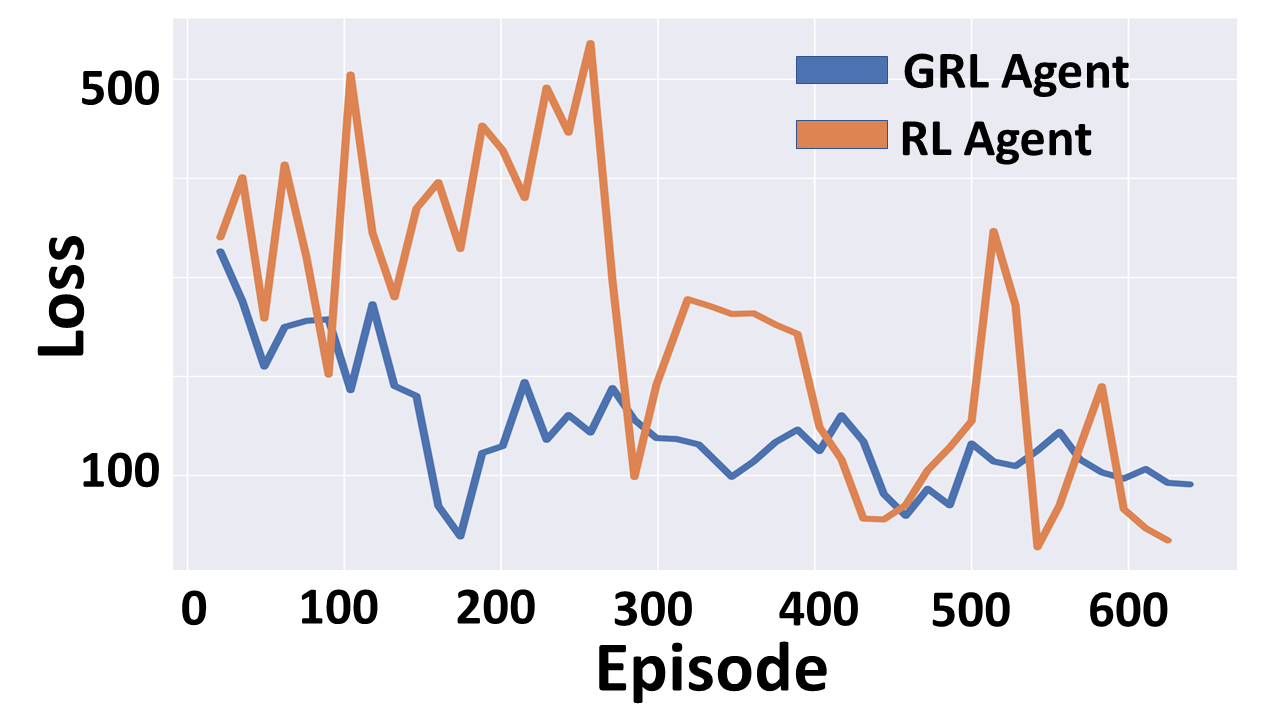}
        % \caption{Loss}
        \label{fig:grlagentnetwork}
    \end{minipage}
    \vspace{-10mm}
    \caption{Reward (left) and Loss (right) plots during training of the RL and GRL Agent for 632 Episodes}
    \label{fig:convergence_plots} 
\end{figure}

% We consider 4 Vtols and 3 ports for the training, all the vehicles are assigned a random schedule at the beginning. The main objective of the agent is to decide the best action for a vehicle considering all the other vehicles status. Figure \ref{} shows the convergence history of graph learning and the figure \ref{} shows the delay time and collision information. As the figure explains the number of collisions is considerably higher in the beginning and as the agent learns to understand the correlation between locations of other agents, the number of collisions over the episodes decreases. The same applies for the number of perfect take-offs and landing.
\vspace{-2mm}
The reward and loss for each agent during training are shown in figure \ref{fig:convergence_plots}. The GRL agent was able to converge, while the RL agent struggled. The GRL agent shows a stable loss decline, while the RL agent's loss exhibits more oscillations, indicating that the GRL agent found it easier to correlate its observations in the state space.

\vspace{-2.1mm}
\subsection{Baseline Comparison}
\vspace{-2.1mm}
\begin{figure}
\centering
\includegraphics[width=.9\linewidth]{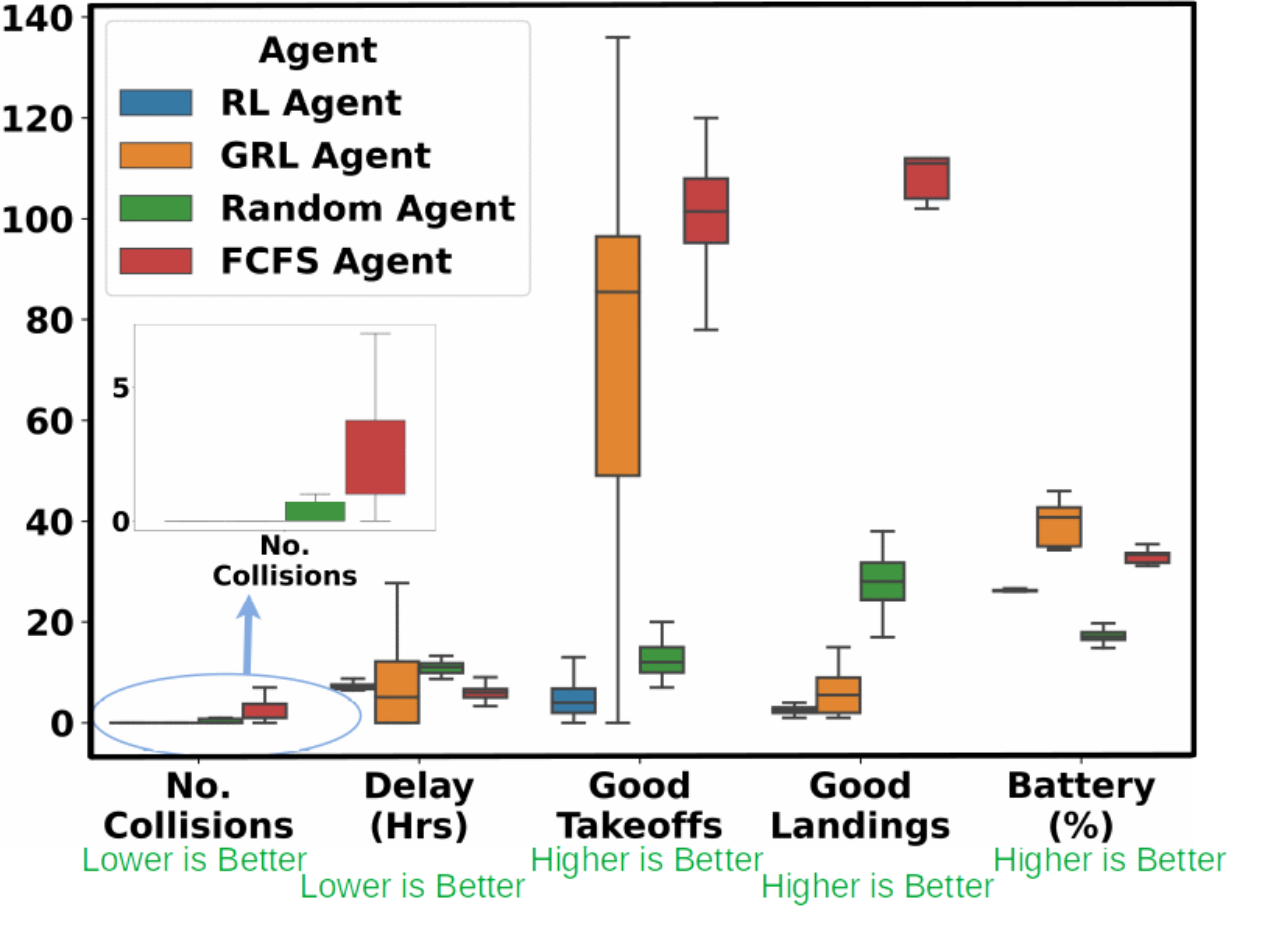}
\vspace{-3mm}
\caption{Comparison of all the reward terms against the baseline methods.}
\label{figs:testresults}
\end{figure}

In order to evaluate the performance of the GRL agent, we compare it with an RL agent with a Multi-Layer Perceptron (MLP) as feature extractors (instead of GNN) trained with the same parameters referenced in section \ref{sec:learning}, a random agent, and a First-Come-First-Served (FCFS) agent (adapted from \cite{andreeva2012study}). The random agent chose random feasible actions from the action space during each decision-making instance. In the FCFS approach, each vehicle ready for landing is put in a queue for recharging (for up to 6 steps), and the vehicle which is ready to take off after recharging is put into another queue. A scheduler commands each vehicle in the queues to execute its action one after the other. Figure \ref{figs:testresults} show the results after testing each agent for 50 episodes (72000 steps). It is important to note the delay depicted is cumulative across all 4 vehicles per episode. The GRL and FCFS agent are observed to have the best performances of the four when it comes to cumulative reward (GRL: $~565 \pm 522.7$, FCFS: $~293 \pm 86$), delay, good takeoffs, and battery management. This is expected, as the GRL agent learned the feature space faster and its goal is to maximize all metrics, while the FCFS agent takes off and lands at set intervals that align closely with each vehicle's schedule and maximize specific metrics (takeoffs, landings, and battery management). As for collisions, the GRL agent has a mean of $~0.3 \pm 1.7$, which is $76\%$ smaller than that of the RL agent and $90\%$ smaller than the FCFS agent, which has means of $~0.86 \pm 4.24$ and $~2.88 \pm 0.66$ respectively. The RL agent does outperform the GRL agent in time scheduling efficiency, with a mean delay of $~7.49 \pm 1.43$ hours per vehicle, which is an $80\%$ lower standard deviation compared to the GRL agent. The random agent struggled to learn delay, collision, and battery metrics.

\vspace{-0.2cm}
\subsection{Analysis of the Safety Coefficient}
\vspace{-1.5mm}
\begin{figure}
\centering
\includegraphics[width=.9\linewidth]{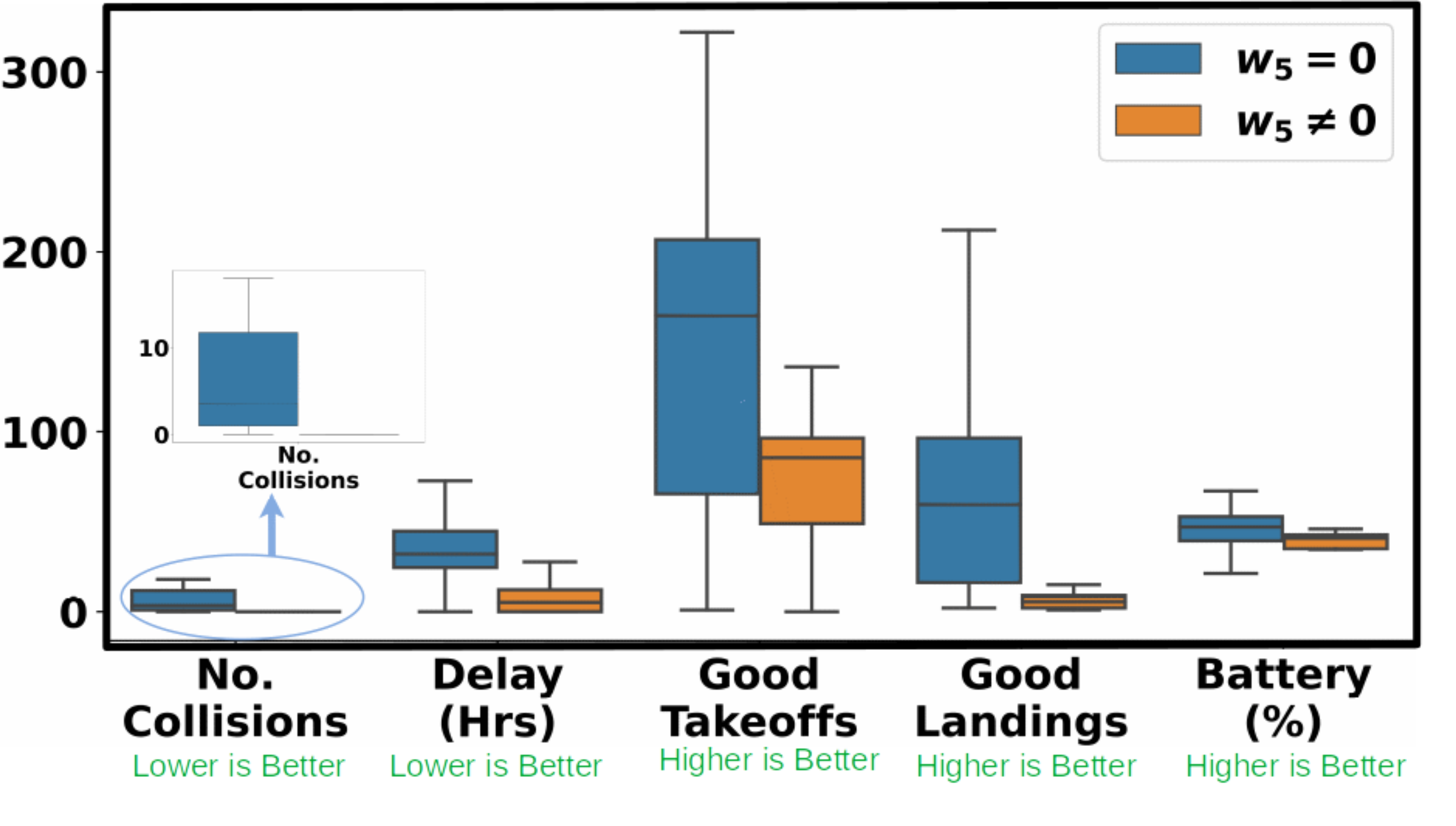}
\caption{Analysis of the effect of the Safety Weight on GRL Agent}
\label{figs:safety_comparison}
\vspace{-2.5mm}
\end{figure}
In this section, we analyze the importance of the safety parameter $\S$. Here, we run the training for 2 different values of $w_5$ in the reward function (eqn. \ref{eq:reward}), $w_5 = 0$
and $w_5 = 2.2$. These models are then evaluated for 50 episodes and the results are shown in Fig. \ref{figs:safety_comparison}. The model trained without a safety weight: {\bf i.)} got into more collisions over the course of an episode; {\bf ii.)} experienced more takeoffs and landings, which led to it outperforming the model with a safety weight and also experiencing more delay. This analysis shows that more training time and variety is required to understand an environment with both delay and safety factors, as the agent has seemingly less difficulty learning one or the other and is forced to compromise when learning both.

\vspace{-3mm}
\section{CONCLUSION}
\label{sec:conclusion}
\vspace{-1.5mm}
In this paper, we proposed a Graph RL (GRL) method to generate a policy that serves as the real-time air traffic control (ATC) system at a vertiport. Specifically, this policy performs real-time scheduling of take-offs and landings of VTOL aircraft, considering safety, delay, and battery level. We formulated the UAM vertiport management problem as an MDP with the state of vertiport and states of aircraft expressed as two separate graphs. We use two GCNs to extract the vertiport and aircraft state information as learnable feature vectors. The training was performed using PPO on a reward function defined as a weighted combination of landing/takeoff delay, number of collisions, and battery consumption, across the vehicles being managed. 
GRL policy performance is compared with three baselines: MLP-based RL agent, Random agent, and First-Come-First-Served. Due to GCN's better feature extraction, GRL outperformed MLP-RL, learning faster and achieving higher episodic rewards. In test episodes, GRL surpassed MLP-RL and the random agent, demonstrating its generalizability. %A parametric study regarding the reward composition indicated that future regulations and market needs (guiding safety, efficiency, delay etc.) could significantly impact/shift the policy performance. 
In the future, simulation-based nonlinear optimization could enhance RL solutions, approximating the optimality gap. Improved draft interactions with nearby aircraft could also boost the effectiveness of the proposed UAM vertiport management approach.
\vspace{-.2cm}
\addtolength{\textheight}{-2cm}   % This command serves to balance the column lengths

%%%%%%%%%%%%%%%%%%%%%%%%%%%%%%%%%%%%%%%%%%%%%%%%%%%%%%%%%%%%%%%%%%%%%%%%%%%%%%%%

%%%%%%%%%%%%%%%%%%%%%%%%%%%%%%%%%%%%%%%%%%%%%%%%%%%%%%%%%%%%%%%%%%%%%%%%%%%%%%%%

%%%%%%%%%%%%%%%%%%%%%%%%%%%%%%%%%%%%%%%%%%%%%%%%%%%%%%%%%%%%%%%%%%%%%%%%%%%%%%%%

%%%%%%%%%%%%%%%%%%%%%%%%%%%%%%%%%%%%%%%%%%%%%%%%%%%%%%%%%%%%%%%%%%%%%%%%%%%%%%%%
% \vspace{-2mm}
% \bibliographystyle{IEEEtran}
% \vspace{-1.2mm}

% \bibliography{iros_2023}

\begin{spacing}{0.91}
\bibliographystyle{IEEEtran}
\bibliography{iros_2023}
\end{spacing}

\end{document}